# Non-magnetic doping induced magnetism in Li doped $SnO_2$ nanoparticles

S. K. Srivastava[1*], P. Lejay[2], A. Hadj-Azzem[2] and G. Bouzerar[2]

[1]Institute of Condensed Matter and Nano-sciences, Université Catholique de Louvain, Louvain-la-Neuve, Belgium-1348
[2]Institut Néel, CNRS Grenoble and Université Joseph Fourier, Grenoble, Cedex 9, France-38042

## Abstract

We address the possibility of non-magnetic doping induced magnetism, in Li doped $SnO_2$ nano-particles. The compounds have been prepared by solid state route at *equilibrium* and were found to be crystallized in single rutile phase. The magnetization measurements have shown that Li-doping induces magnetism in $SnO_2$ for a particular range of Li concentration. However, for other Li concentrations, including pure $SnO_2$, the samples exhibit diamagnetism. To investigate the possible origin of the induced magnetism, we have studied the variation of the magnetization as a function of the average nano-particle radius. Possible scenarios for the appearance of magnetism in these compounds are discussed.

---

[*] **E-mail:** sandeep.srivastava@uclouvain.be

# I. INTRODUCTION

Over the last decade, transition metal (TM) doped ferromagnetic semiconductors have drawn a considerable interest as demonstrated by the huge existing literature [1–4]. The main goal of studying TM ferromagnetism is to incorporate such material in devices where the spin degree of freedom would be utilized to carry information. This would lead to a reduction in the power consumption, as well as allow non-volatile storage and data processing at or beyond room temperature. However, there have been controversial experimental reports on the magnetic properties of TM doped oxides [5-6]. Despite TM doped ferromagnetism, unexpected ferromagnetism known as $d^0$ has been reported or predicted in several oxides like $HfO_2$, CaO, ZnO, $ZrO_2$ and even in $CaB_6$ [7-11]. Thus the *$d^0$* or intrinsic ferromagnetism was believed to provide an alternative pathway to TM induced ferromagnetism. However, the origin of the ferromagnetism in such materials is still controversial for most of the cases. From *ab initio* studies, it has been shown that point defects such as cation vacancies could be the origin of the magnetism in some of these materials such as; $HfO_2$ or CaO [8, 12-14]. In a recent model, which includes disorder and electron-electron correlation effects on equal footing, it has been proposed that high Curie temperatures could be reached in oxides such as $A_{1-x}B_xO_2$ (A=Ti, Zr, or Hf) where B is a monovalent cation of the group 1A [15]. The non-magnetic dopant induces local moments on the neighboring oxygen atoms which then interact with extended ferromagnetic exchange couplings. These findings have been followed by many *ab initio* studies and they have predicted high $T_C$ ferromagnetism in several oxides such as; K-$SnO_2$ [16], Mg–$SnO_2$ [17], anatase Li-$TiO_2$ [18], rutile K–$TiO_2$ [19], V-$TiO_2$ [20], and K–$ZrO_2$ [11, 19]. From experimental point of view, such non-magnetic doping induced magnetism has been observed in several oxides such as; alkali metal doped ZnO [21-23]; Cu doped $TiO_2$ prepared in thin film form [24, 25]**,** carbon-doped $TiO_2$ prepared by solid state route [26], K: $SnO_2$ [27] and K: $TiO_2$ [28].

Because of their potential interest for spintronic devices, the search for suitable oxides, appropriate non-magnetic dopants and optimal preparation procedure to obtain room temperature ferromagnetism became really intense. But, in most of the cases (for example, thin film); preparation of materials is not yet very well controlled. Therefore, it is imperative to prepare bulk materials at equilibrium conditions, which will intrinsically diminish the uncertainties and inaccuracies in characterization. $SnO_2$ is a wide band-gap material with a band gap of about 3.6 eV, used as a transparent conduction electrode in flat panel display and solar cells [29]. It has a rutile structure with distorted octahedral coordination. Experimentally, the possibility of magnetism in Li doped $SnO_2$ has not been investigated yet. In this manuscript, we have investigated the possibility of non-magnetic doping induced magnetism in Li doped $SnO_2$.

## II. EXPERIMENTAL METHODS

We have prepared $Sn_{1-x}Li_xO_2$ compounds ($0 \leq x \leq 0.12$) by standard solid state route method by using high-purity $SnO_2$ (purity, 99.996%) and $Li_2CO_3$ (99.998 %) compounds. The final annealing in pellet form was carried out at $500^0C$ for 30 hours in air. Slow scan powder X-Ray diffraction (XRD) patterns were collected by using Philips XRD machine with $CuK_\alpha$ radiation. The recording of microstructure images have been carried out by using Zeiss-Ultra Scanning Electron Microscope (SEM) equipped with an energy dispersive spectrometer (EDS). Magnetization measurements as a function of magnetic field (H) were carried out using a commercial SQUID magnetometer (Quantum Design, MPMS).

## III. RESULTS AND DISCUSSIONS

The X-Ray diffraction patterns for $Sn_{1-x}Li_xO_2$ compounds are shown in Figure 1. All the diffraction's peaks could be indexed on the basis of the tetragonal rutile type-structure. No extra diffraction peaks were detected showing that no crystalline parasitic phases are present in the samples within the limit of XRD. These XRD patterns were refined with the help of the fullprof program by the Rietveld refinement technique [30]. A typical XRD pattern along with refinement is shown in Figure 2 for $Sn_{0.97}Li_{0.03}O_2$ compound. We can clearly see that the experimental X-ray peaks are perfectly matching with power-diffraction software generated x-ray peak. The lattice parameters for pure $SnO_2$ are found to be a = b = 4.7385 Å and c = 3.1871 Å, and are comparable to those reported by Duan *et al*. [31]. However, we do not observe any significant change in the lattice parameters. This can be understood by the fact that $Sn^{4+}$ (0.71Å) and $Li^+$ (0.68 Å) have very similar ionic radii. To understand the microstructure, we have performed observations by SEM. One typical SEM image of $Sn_{0.91}Li_{0.09}O_2$ compound prepared at 500˚C is shown in Figure 3 (a). The morphology of all samples was found to be uniform and it showed conglomerates of nanometric grains. Energy dispersive spectroscopy was carried out to check the presence of any unwanted magnetic impurity. EDS analyses confirm that there is no trace of any kind of magnetic impurity in the compounds within the instrumental limit as shown in the Figure 4. Thus, from the X-ray and SEM analyses, we can conclude that all compounds have been crystallized to single phase of tetragonal rutile type-structure of $SnO_2$.

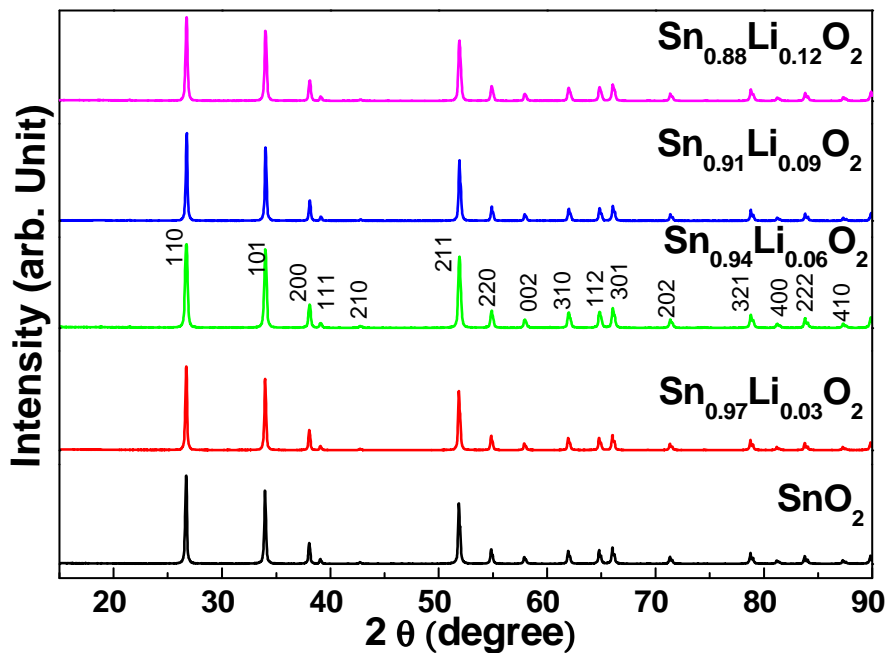

**Figure 1:** X-Ray Diffraction patterns of $Sn_{1-x}Li_xO_2$ ($0 \leq x \leq 0.12$) compounds.

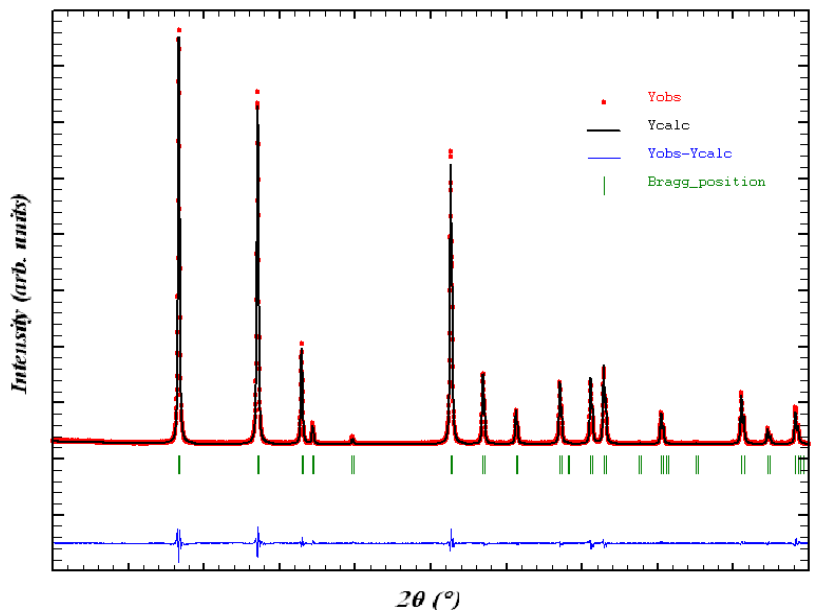

**Figure 2:** Refinement of XRD patterns for $Sn_{0.97}Li_{0.03}O_2$ compound

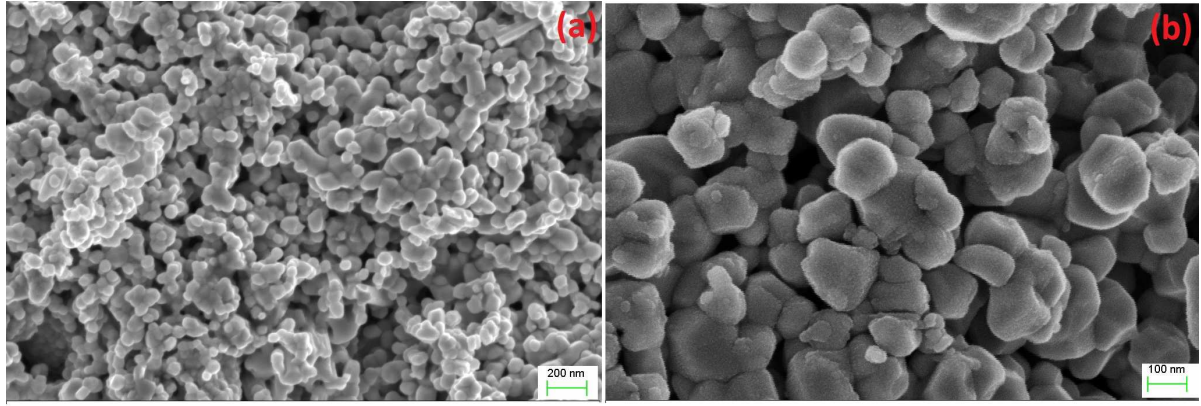

**Figure 3:** SEM micrograph of $Sn_{0.91}Li_{0.09}O_2$ compound **(a)** prepared at 500 $^0$C **(b)** prepared at 800 $^0$C.

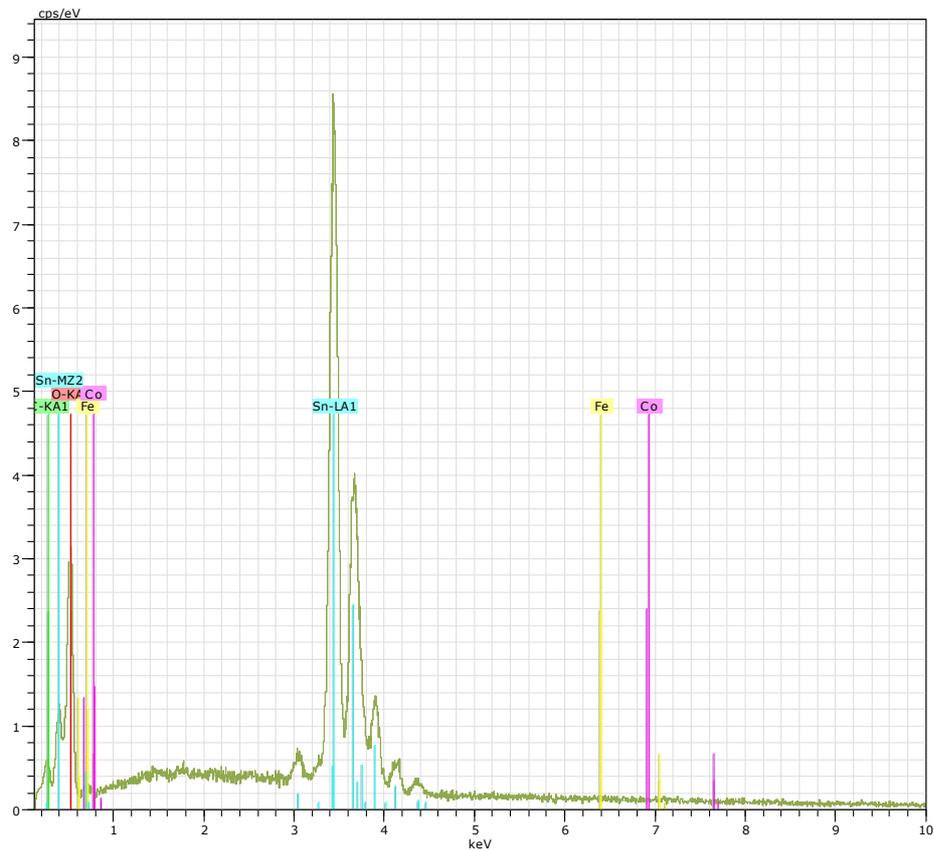

**Figure 4:** Energy dispersive spectroscopy of $Sn_{0.91}Li_{0.09}O_2$ compound prepared at 500 $^0$C. The olive color curve represents all detected peaks. The expected positions for magnetic impurities like Co and Fe are marked in the curve. It confirms that there is no trace of any magnetic impurity in these compounds.

The magnetic measurements for all samples were done with SQUID magnetometer with utmost care and repeated in triplicate with different pieces of samples to guarantee the reproducibility of results. The magnetic properties of all the starting compounds; $SnO_2$ and $Li_2CO_3$ were also checked and they clearly exhibit diamagnetic behavior. The M-H measurements for $Sn_{1-x}Li_xO_2$ compounds show that both, pure $SnO_2$ and 3% Li doped compound exhibit a clear diamagnetic behavior at 3 K, as illustrated in Figure 5 (a). However, the 6% and 9% doped compounds are surprisingly found to be magnetic. Their magnetization is found to increase with Li concentration, as shown in Figure 5 (a). It approaches saturation for the 9% Li doped compound with a magnetic moment of 0.0022 emu/gm at 3 K and 5 Tesla field. However, for a larger concentration x=0.12, it again exhibits diamagnetic behavior. Thus, to summarize, Li doping in $SnO_2$ leads to magnetic moment formation only for a small window of Li-concentration ($0.03 \leq x \leq 0.12$). Nevertheless, within this concentration range, the compounds exhibit weak paramagnetism, but no long-range ferromagnetic order.

To understand the origin of the observed magnetism in these compounds, we have focused our attention on the 9% doped compound. Several different samples were obtained by changing the annealing temperature. The samples were systematically annealed for 20 hours after pre-sintering them at $300^0$C for about 30 hours in the air. The annealing temperature ranges from 400°C to 800 °C. The XRD patterns for all samples have been refined by using the tetragonal rutile type-structure. All compounds were found to be in a single-phase form. Two typical microstructural images obtained from SEM are shown in Figure 3a & 3b for samples prepared at 500 and $800^0$C. The morphology obtained from SEM was found to be quite uniform and it shows conglomerates of nano-sized grains. The average particles size of the grains was obtained by analyzing several frames of images. The error of measurement was of the order of 2 nm. The average particle size (*D*) obtained from the above analysis was 60, 90, 130, 160, 190 nm for the compounds prepared at 400°C, 500°C, 600°C, 700°C and 800°C respectively. The M-H measurements for the series of $Sn_{0.91}Li_{0.09}O_2$ compounds, prepared at various temperatures are found to be quite interesting. The M-H measurement shows that the sample prepared at 400°C exhibits a diamagnetic behavior at 3 K, as shown in Figure 5 (b). However, the compounds prepared at higher temperature are found to be magnetic. Their magnetization increases sharply with their average radii. The magnetization of the sample prepared at 500°C is almost seven times to that of the compound prepared at 800°C, as seen in Figure 5 (c).

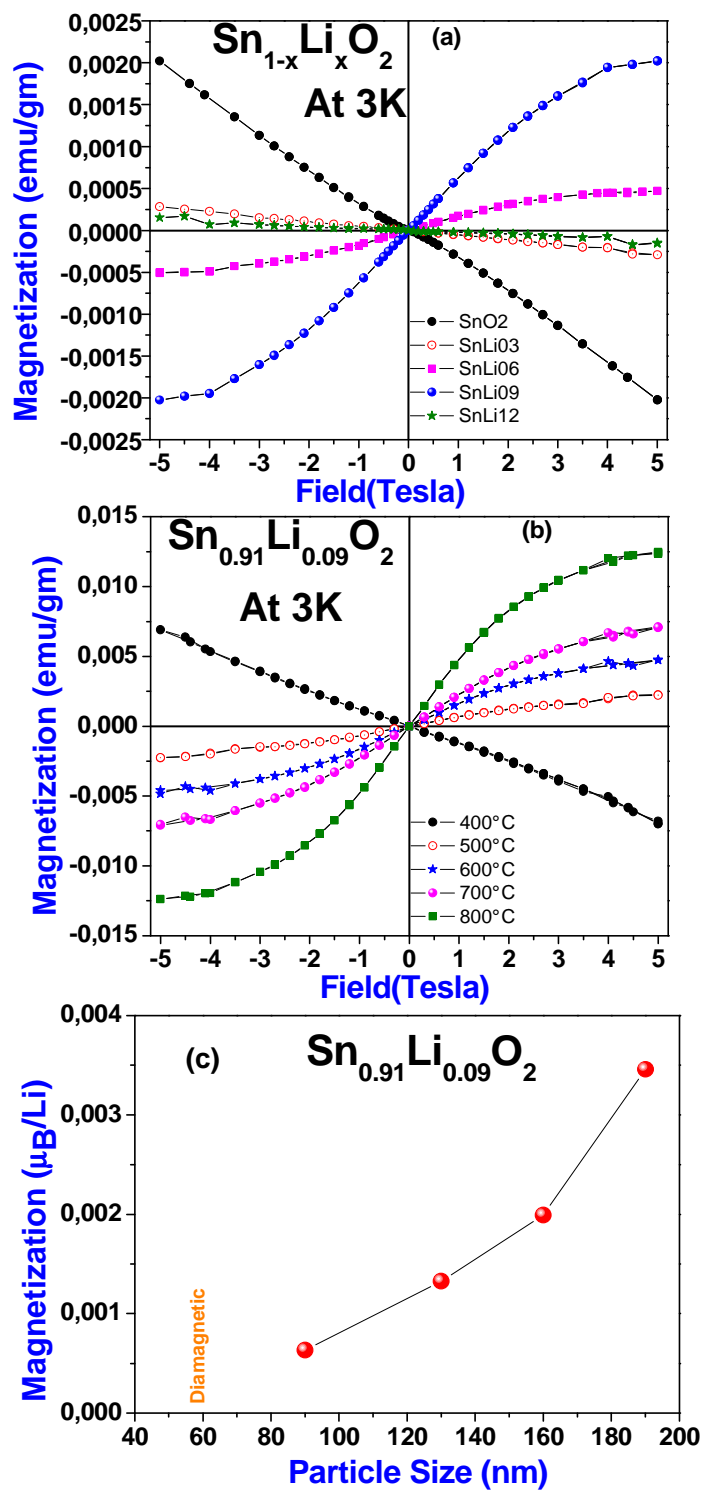

**Figure 5: (a)** M-H loops recorded at 3 K for $Sn_{1-x}Li_xO_2$ compounds prepared at 500°C **(b)** M-H loops recorded at 3 K for $Sn_{0.91}Li_{0.09}O_2$ compounds prepared at various temperatures. (c) Variation of magnetic moment (at 3K, 5 Tesla) with average particle size for $Sn_{0.91}Li_{0.09}O_2$ compounds prepared at various temperatures.

Now, let us discuss the possible origin of magnetism in these Li doped $SnO_2$ compounds in the light of existing first-principles calculations. We discuss three possible scenarios to explain the experimentally observed magnetism.

(i) *Bulk magnetism*: In the case of a direct cationic substitution, theoretical studies [11, 15] have demonstrated that three physical parameters are essential to explain induced $d^0$ magnetism: (i) the position of the induced impurity band which should be located near the top of the valence band, (ii) the density of carrier per defect, and (iii) the electron-electron correlations. The substitution of $Sn^{4+}$ by $Li^+$ in pure $SnO_2$ provides three holes in the present case. In a recent first-principles calculation for Li doped $SnO_2$, it has been shown that Li substitution induces magnetism in $SnO_2$ [32]. A large magnetic moment of $3\mu_B$ has been obtained. The low-lying s orbitals of Li are spin-polarized and strongly hybridized with the p orbitals of O. The Fermi energy is mainly dominated by the p orbitals of O, which indicates that magnetism is mainly induced in the p orbitals, localized at the O atom. Indeed, the oxygen atoms surrounding the Li ion provide the dominant contribution to the total magnetic moment. Moreover, they have shown that there is a very small (negligible) induced magnetic moment at the Li site, which suggests that Li behaves as a spin polarizer in $SnO_2$ [32]. These results are consistent with the general picture provided in ref. [15] and discussed in ref. [11] in the case of K and Na doped $ZrO_2$. From our experimental results, we have found much smaller moments, (see Figure 5c). Indeed, the moment found for the largest nano-particle was of the order of 0.0035 $\mu_B$/Li.

*(ii) Native defect induced Magnetism*: While preparing the samples, the increase in preparation temperature causes an increase in particle size; this may not be the only change that occurs. In particular, defect formation/ vacancies in the bulk may be temperature dependent; and it may be those defects that are responsible for the magnetism. These defects could be oxygen vacancies or cationic vacancies, for example refer [8, 12, 14, 15]. From *ab initio* based studies (which have their own limitations), oxygen vacancies do not lead to magnetic moment formation in most of the cases. However cationic vacancies lead to large moments. Thus, it would be of interest to address these possibilities theoretically in $SnO_2$. From our data, we do not believe that this scenario is likely.

*(iii) Surface induced Magnetism:* If the observed moment was surface-induced, one would normally expect the magnetization (M) to be inversely proportional to the average particle radius ($R_{avg}$). However, we observe an increase of M with respect to $R_{avg}$ from our experimental data. Our data seems to contradict this scenario. However, if one takes into account the fact that the particle sizes obey a Gaussian distribution with a certain width and that below a certain radius $R_C$, the nano-particles remain diamagnetic. Thereafter, one would get first an increase as a function of R and then a 1/R behavior, only for R>> $R_C$. Thus, this scenario appears to be a possible explanation. The calculated moment per Li(s) is found to vary from 0.1 to 2.6 $\mu_B$ for the largest radius. Note that, similar results showing the crucial role played by the surface has been reported from *ab initio* studies. More precisely, it has been shown that C induces a magnetic moment at the surface only in the C: $SnO_2$ compound [33]. The origin of the moment was attributed to surface bonding. Indeed, at the surface, the numbers of bonds are reduced leading to unpaired electrons that follow the Hund's rule to minimize the Coulomb repulsion. As a result, one finds a large induced moment of $2\mu_B/C$.

## IV. CONCLUSION

To conclude, we have prepared $Sn_{1-x}Li_xO_2$ (x=0-0.12) compounds by solid state route method. The X-ray diffraction and the detailed micro structural analyses provide the evidence of single rutile phase. We have shown that 6% and 9% Li-doped compounds exhibit a magnetic phase at 3 K. However, other Li doped compounds, including $SnO_2$ are diamagnetic. We have also studied the effect of annealing temperature on the magnetization in order to understand the observed magnetism and we have speculated the possible origin of observed magnetism. However, further advanced study is required to know the exact origin of magnetism in these samples. The experimental tools such as X-Ray magnetic circular dichroism can help to probe local magnetism. We hope that our results will stimulate further experimental studies in these compounds. It would be exciting to find out whether other cations with different valence such as $Zn^{2+}$ or $Mg^{2+}$ with respective ionic radii 0.74 Å and 0.65 Å, could also induce a magnetic moment in $SnO_2$ nano-particles.